\documentclass[aps,prl,twocolumn,showpacs]{revtex4}
\usepackage{bm}
\usepackage{amsmath}
\usepackage{amssymb}
\usepackage{graphicx}

\newcommand{\beq}{\begin{equation}}
\newcommand{\eeq}{\end{equation}}
\newcommand{\beqar}{\begin{eqnarray*}}
\newcommand{\eeqar}{\end{eqnarray*}}

\newcommand{\rtarr}{\rightarrow}

\newcommand{\om}{\omega}

\newcommand{\la}{\lambda}

\newcommand{\eps}{\varepsilon}
\newcommand{\ka}{\varkappa}

\begin{document}

\title{Electron screening and excitonic condensation in double-layer graphene
systems}
\author{Maxim Yu. Kharitonov$^{1}$ and Konstantin B. Efetov$^{1,2}$}
\affiliation{$^{1}$ Theoretische Physik III, Ruhr-Universit\"{a}t Bochum, Germany\\
$^{2}$L.D. Landau Institute for Theoretical Physics, Moscow, Russia}
\date{\today}

\begin{abstract}
We theoretically investigate the possibility of excitonic
condensation in a system of two graphene monolayers separated by
an insulator, in which electrons and holes in the layers are
induced by external gates. In contrast to the recent studies of
this system, we take into account the screening of the interlayer
Coulomb interaction by the carriers in the layers, and this
drastically changes the result. Due to a large number of electron
species in the system (two projections of spin, two valleys, and
two layers) and to the suppression of backscattering in graphene,
the maximum possible strength of the screened Coulomb interaction
appears to be quite small making the weak-coupling treatment
applicable. We calculate the mean-field transition temperature for
a clean system and demonstrate that its highest possible value
$T_c^\text{max}\sim 10^{-7}\epsilon_F\lesssim 1\,\text{mK}$ is
extremely small ($\epsilon_F$ is the Fermi energy). In addition,
any sufficiently short-range disorder with the scattering time
$\tau \lesssim \hbar /T_c^\text{max}$ would suppress the
condensate completely. Our findings renders experimental
observation of excitonic condensation in the above setup
improbable even at very low temperatures.
\end{abstract}

\maketitle

\section{Introduction and main result}
The possibility of excitonic condensation (EC) in metallic systems
was originally proposed by Keldysh and Kopaev~\cite{KK} for
semimetals with overlapping conduction and valence bands. They
have shown that the attractive Coulomb interaction between
electrons and holes leads to an instability towards formation of
bound electron-hole pairs analogous to the Cooper instability in
superconductors. Somewhat later, it was suggested~\cite{LY}
that EC could be realized in a double-layer system of spatially
separated electrons and holes. Experimental efforts towards the
observation of EC were mainly concentrated on
semiconductor double quantum well
systems~\cite{exp1,exp2,exp3,exp4} and experimental data speak in
favor of the existence of EC in electron-hole
bilayers~\cite{exp1,exp2,exp3} and electron-electron bilayers in
the quantum Hall regime~\cite{exp4}.

Since the carrier density in graphene, including its polarity, can
effectively be controlled by various means, graphene-based systems
may also seem attractive for the realization of EC. Indeed,
several ideas on how one could obtain EC in graphene have been
suggested recently.
One possible way to create interacting electrons and holes is to
apply a strong in-plane magnetic field to a single layer of
graphene~\cite{AKT}. Such a magnetic field acts on the spins of
the carriers only, and the Zeeman splitting creates electrons with
one spin polarization and holes with the opposite polarization in
an initially neutral sample. A detailed theory of EC in such a
setup has been developed in Ref.~\cite{AKT}.

\begin{figure}[tbp]
\includegraphics[width=.40\textwidth]{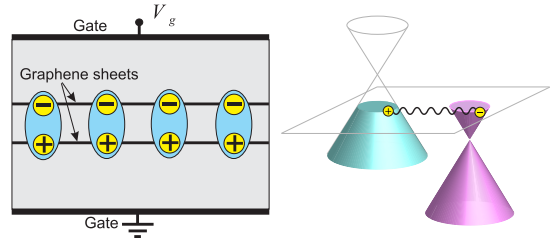}
\caption{ Excitonic condensate in a system of two spatially
separated graphene layers. Electrons and holes in the layers are
induced by applying the external gate voltage.} \label{fig:system}
\end{figure}

A double-layer graphene system (Fig.~\ref{fig:system}) as a
candidate for the observation of EC was proposed recently in
Refs.~\cite{MD1,MD2,ZJ}.  If two graphene layers are separated by
an insulator, electrons in one layer and holes in the other can be
obtained by applying external gate voltage. Relatively high values
of the Fermi energy $ \epsilon _{F}\sim 0.3\,\text{eV}$ that can
be achieved in graphene by using gates~\cite{graphenenature} are
an obvious advantage, since $\epsilon _{F}$ serves as a
high-energy scale of the effect in such a setup. Solving the gap
equations numerically, the authors of Refs.~\cite{MD2,ZJ} provided
an estimate $T_{BKT}\sim 0.1\epsilon _{F}$ for the critical
temperature of the Berezinski-Kosterlitz-Thouless (BKT) transition
and argued that $T_{BKT}$ could thus reach room temperatures.

However, in the analysis of Refs.~\cite{MD2,ZJ}, the screening of
the Coulomb interactions by the carriers in the graphene layers
was not taken into account. Clearly, the two-dimensional screening
cloud formed around a probe charge in a graphene sheet screens the
field of the charge both in the off-plane and in-plane directions,
although not identically. Therefore, screening affects not only
the intralayer but also the interlayer Coulomb interaction in the
double-layer setup.

In this paper, we demonstrate that taking screening into account
is essential  as it drastically reduces the transition temperature
compared to the estimate obtained in Refs~\cite{MD2,ZJ} neglecting
screening. In fact, screening sets the upper bound for the
interaction strength, yielding for the maximum possible value of
the dimensionless coupling constant
\begin{equation}
\lambda ^{\text{max}}\ =\frac{1}{N}=\frac{1}{8}.  \label{eq:lamax}
\end{equation}%
Here $N=N_{s}N_{v}N_{l}=2^{3}=8$ is the total number of electron
species in the system originating from two projections of spin
($N_s=2$), two valleys ($N_v=2$), and two layers ($N_l=2$).
Moreover, the chiral nature of quasiparticles in graphene leads to
the suppression of backscattering. Consequently, the maximum
interaction strength $\lambda _{c}^{\text{max}}$ that determines
the transition temperature appears to be actually two times
smaller
than $\lambda ^{\text{max}}$, 
\begin{equation}
\lambda _{c}^{\text{max}}=\frac{\lambda ^{\text{max}}}{2}=\frac{1}{16}.
\label{eq:lamaxBCS}
\end{equation}

As follows from Eqs.~(\ref{eq:lamax}) and (\ref{eq:lamaxBCS}), the
large number of electron species and suppression of backscattering
make the maximum possible value $\lambda _{c}^{\text{max}}$ of the
interaction strength numerically quite small. This justifies the
applicability of the weak-coupling BCS approach to the problem,
since 1/16 can safely be considered as a small parameter. As a
result, for the highest possible value of the mean-field
transition temperature we obtain
\begin{equation}
T_{c}^{\text{max}}\approx \exp (-1/\lambda _{c}^{\text{max}})\,\epsilon
_{F}=\exp (-16)\,\epsilon _{F}\approx 10^{-7}\epsilon _{F}.  \label{eq:Tcmax}
\end{equation}%
This is the {\em highest possible} value of the critical
temperature of the excitonic condensation that could be achieved
in a perfectly clean double-layer graphene system. Entering
Eq.~(\ref{eq:Tcmax}) as the exponent, the large value of the
inverse interaction strength $1/\lambda _{c}^{\text{max} }=16$
results in a drastic reduction of the transition temperature. In
order to achieve the maximum value (\ref{eq:Tcmax}), the
interlayer distance $d$ must be much smaller than the Debye
screening length $\ka^{-1}$,
\begin{equation}
2 \varkappa d\ll 1.  \label{eq:smalld}
\end{equation}%
The most optimistic estimate would thus be $T_{c}^{\text{max}}\sim
1\,\text{mK}$ for $\epsilon _{F}\sim 0.3\,\text{eV}$ and would
require $d\lesssim 0.2\,\text{nm}$.

\section{Calculations}

We  now present the details of derivation of Eq.~(\ref{eq:Tcmax}).
The bare strength of Coulomb interactions in graphene is not that
small. For SiO$_2$ used as an insulator embedding graphene sheets,
typical values of the dimensionless coupling constant are
$r_s=e^2/(\eps v) \sim 1$ ($\eps$ is the dielectric constant of
the insulator and $v$ is the velocity of the Dirac spectrum, we
set $\hbar=1$ throughout this section). This questions the
applicability of the weak-coupling approach to the problem of EC,
suggesting, at the same time, that the transition temperature
could be quite high~\cite{MD2,ZJ}.

It is known, however, that in a fermionic system with a large
number $N \gg 1$ of independent fermionic species the interactions
are significantly weakened. Physically, large $N$ makes screening
very effective, since {\em all} $N$ species participate in the
screening of interactions between fermions of {\em each}
particular species. Screening reduces the coupling constant from
$r_s$ to the value $1/N \ll 1$: $r_s \rtarr 1/N$. Effectively, the
system becomes weakly interacting, despite of the fact that the
bare Coulomb interactions may be not weak ($r_s \gtrsim 1$).

The  large-$N$ approximation was  already  used for a single-layer
graphene~\cite{Son,AKT,FA} before. In a single layer, the number
of species is equal to $N_1=N_s N_v =4$ due to two projections of
spin ($N_s=2$) and two valleys
($N_v=2$). 
This value is not exceptionally large, but does give hope that the
large-$N$ approach adequately describes graphene physics. In a
double-layer system the situation is better: since each electron
can belong to either one of the layers,  one has an additional
``which-layer'' degree of freedom ($N_l=2$)  making the total
number of species $N=N_s N_v N_l=8$. It would already be quite
reasonable to treat $N=8$ as a large parameter. Therefore,
large-$N$ approximation seems to be particularly suitable for a
double-layer graphene system and is expected to provide good
quantitative predictions.

Below we employ the large-$N$ approach to the double-layer
graphene system (Fig.~\ref{fig:system}) treating $N=8$ as a large
parameter and calculate the mean-field  critical temperature $T_c$
of EC. The calculations follow closely those of Ref.~\cite{AKT}.
Of course, the mean-field treatment of two-dimensional systems is
not necessarily a good one due to strong thermal fluctuations. It
is, however, sufficient for our purposes, as our main goal is to
demonstrate that already the mean-field critical temperature is
extremely low. The temperature $T_{BKT}$ of the actual BKT
transition can only be lower than the mean-field $T_c$ we
calculate here.

Within the large-$N$ approximation, the diagrammatic series for
the effective interaction between the electrons is identical to
that of the random-phase approximation (RPA), which describes
linear screening.
%
For the problem at hand, the relevant transfer momenta $q$ are in
the range $0\leq q \leq 2 p_F$, with $p_F=\epsilon_F/v$ the Fermi
momentum, and one has to use the exact expression for the
polarization operator (and not its limit form for $q\ll p_F$).
However, the static polarization operator
 in graphene~\cite{PO} does not
depend on momentum  at all in this range 
and equals $\Pi(\om=0,q)=N_s N_v \nu$, where $\nu=\epsilon_F/(2\pi
v^{2})$ is the density of states per one valley and one spin. As a
result, for the screened interlayer Coulomb interaction in
the momentum space 
we obtain
    \begin{equation}
        V(q)=\frac{2\pi e_{\ast }^{2}\exp(-qd)}
            {q+2\varkappa+\varkappa^2[1-\exp (-qd)]/q}, \mbox{ } q
            \leq 2 p_F.
        \label{eq:V}
    \end{equation}
In Eq.~(\ref{eq:V}), $q$ is the absolute value of the in-plane
two-dimensional wave vector, $d$ is the distance between the
layers, $e_*$ is the effective electron charge screened by the
insulator embedding graphene sheets, $e_*^{2}=e^{2}/\varepsilon $,
and $\varkappa=2\pi N_{s}N_{v}e_{\ast }^{2}\nu$ is the inverse
Debye screening length in each layer. We assume the same Fermi
momenta $p_F$ of electrons and holes (this can be achieved by
tuning  the gate voltage), since any difference between them would
be suppressing the condensate in a way Zeeman splitting suppresses
s-wave superconductivity.

\begin{figure}[tbp]
\includegraphics[width=.35\textwidth]{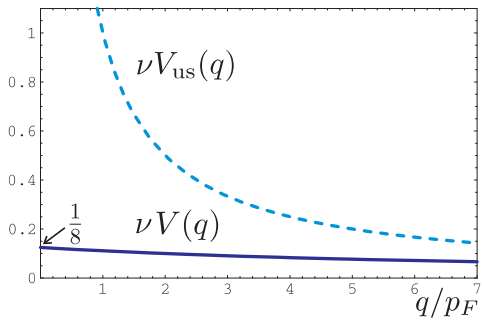}
\caption{ The screened $V(q)$ [Eq.~(\protect\ref{eq:V}), solid
line] and unscreened $V_\text{us}(q)$ [Eq.~(\protect\ref{eq:Vus}),
dashed line] interlayer Coulomb interaction, responsible for the
excitonic instability. The values $d=0$ and $r_s =e^2/(\eps v)=1$
were used, $\nu$ is the density of states. At relevant momenta
$q\sim p_F$, the unscreened interaction potential overestimates
the actual screened one by about $ 10$ times. The
screened Coulomb interaction reaches its maximum at $q=0$, the universal value (%
\protect\ref{eq:lamax}) of which is achieved for $2 \varkappa d
\ll 1$ [Eq.~(\protect\ref{eq:smalld})]. } \label{fig:V}
\end{figure}

The screened Coulomb interaction (\ref{eq:V}) is a decreasing
function of $q$ (Fig.~\ref{fig:V}) and reaches its maximum at
$q=0$,
\begin{equation}
V(q=0)=\frac{2\pi e_{\ast }^{2}}{2\varkappa +\varkappa ^{2}d}.  \label{eq:V0}
\end{equation}%
The maximum of Eq.~(\ref{eq:V0}) is achieved, if the distance $d$
between the layers is smaller than the Debye
radius~[Eq.~(\ref{eq:smalld})], and equals
\begin{equation}
V^{\text{max}}(q=0)=\frac{1}{2N_{s}N_{v}\nu }.  \label{eq:V0max}
\end{equation}%
The factor  $2$ that enters the denominator in
Eq.~(\ref{eq:V0max}) is due to ``which layer'' degree of freedom
($N_l=2$), since each carrier can belong to either one of the
layers. Equation~(\ref{eq:V0max}) leads to Eq.~(\ref{eq:lamax})
for the maximum value of the dimensionless coupling constant
$\lambda ^{\text{max}}=\nu V^{\text{max}}(q=0)$.

In contrast to the above calculation, in the analysis of
Refs.~\cite{MD2,ZJ} the unscreened form
    \begin{equation}
        V_{\text{us}}(q)=2\pi e_{\ast }^{2}\exp (-qd)/q
        \label{eq:Vus}
    \end{equation}%
of the Coulomb interaction $V(q)$ [Eq.~(\ref{eq:V})] was used, see
Eq.~(4) in Ref.~\cite{MD2} and inline formulas before Eq.~(2) in
Ref.~\cite{ZJ}. As seen from Eqs.~(\ref{eq:V}) and (\ref{eq:Vus})
and Fig.~\ref{fig:V},  the unscreened form $V_\text{us}(q)$ is
valid for $q\gg 2\varkappa = N r_s p_F$, but significantly
overestimates the actual screened interaction $V(q)$ for relevant
momenta $q \leq 2 p_F$. For the value $r_s = 1$ used in
Ref.~\cite{MD2} and typical for SiO$_{2}$ as an insulator, one
obtains $V_{\text{us}}(p_{F})\approx 9\,V(p_{F})$. Using the
unscreened form of the Coulomb interaction in Refs.~\cite{MD2,ZJ}
resulted in the estimate $T_{BKT} \sim 0.1 \epsilon_F$ and, as it
appears, led an overestimation of $ T_{BKT}$ by a factor $10^{6}$,
see Eq.~(\ref{eq:Tcmax}).

In order to obtain the mean-field transition temperature $T_c$, we
derive the linearized gap equation for the order parameter
\begin{equation}
\hat{\Delta}(\mathbf{r}-\mathbf{r}^{\prime })=V(\mathbf{r}-\mathbf{r}%
^{\prime })\langle \hat{\phi}_{e}(\mathbf{r})\hat{\phi}_{h}^{\dagger }(%
\mathbf{r}^{\prime })\rangle.
\end{equation}%
Here, $V(\mathbf{r}-\mathbf{r}')$ is the interaction~(\ref{eq:V})
in the coordinate space and $\hat{\phi}_{e,h}$ are the Dirac
spinor fields of electrons and holes in the graphene sheets. The
matrix structure of the order parameter
in the sublattice space is predetermined by chirality, but can be
arbitrary in the valley and spin spaces. Using the standard BCS
approach, we arrive at the linearized gap equation
\begin{equation}
\hat{\Delta}(\mathbf{n})=\nu \ln \frac{\epsilon _{F}}{T}\int \frac{\text{d}%
\mathbf{n}^{\prime }}{2\pi }V(p_{F}|\mathbf{n}-\mathbf{n}^{\prime }|)\hat{%
\mathcal{P}}(\mathbf{n}^{\prime })\hat{\Delta}(\mathbf{n}^{\prime })\hat{%
\mathcal{P}}(-\mathbf{n}^{\prime }),  \label{eq:De}
\end{equation}%
where the two-dimensional unit vectors $\mathbf{n}$ and $\mathbf{n}^{\prime }
$ represent the direction of the electron momentum and $%
\hat{\mathcal{P}}(\mathbf{n})=(1+\mbox{\boldmath{$\tau$}}\mathbf{n})/2$,
with $\tau _{x}$ and $\tau _{y}$  the Pauli matrices in the
sublattice space. The value of temperature $T$, at which a nonzero
solution $\hat{\Delta}(\mathbf{n})$ to Eq.~(\ref{eq:De}) appears,
determines $T_c$. Solving Eq.~(\ref{eq:De}), we obtain
    \begin{equation}
    T_{c}\approx \exp (-1/\lambda _{c})\, \epsilon _{F},
    \label{eq:Tc}
    \end{equation}%
where
    \begin{equation}
    \lambda _{c}=\nu \int_{-\pi }^{\pi }\frac{\text{d}\theta }{2\pi }V\left(
    2p_{F}\sin \frac{\theta }{2}\right) \frac{1+\cos \theta }{2}.  \label{eq:la}
    \end{equation}%
The exact numerical value $\sim 1$ of the prefactor  in
Eq.~(\ref{eq:Tc}) cannot be obtained within the logarithmic
accuracy of the mean-field approach. The form of
Eqs.~(\ref{eq:Tc}) and (\ref{eq:la}) and the matrix structure of
the solution $\hat{\Delta}(\mathbf{n})$ in the sublattice space
are identical to those in Ref.~\cite{AKT} [see Eqs.~(5.23)-(5.31)
therein]. At the same time,  the form (\ref{eq:V}) of the
interaction $V\left( q\right)$ is different here.

The maximum possible value $\lambda _{c}^{\text{max}}$
[Eq.~(\ref{eq:lamaxBCS})] of the interaction constant $\lambda
_{c}$ [Eq.~(\ref{eq:la})] and, thus, the highest possible
transition temperature $T_{c}^{\text{max}},$
[Eq.~(\ref{eq:Tcmax})], are obtained by inserting
Eq.~(\ref{eq:V0max}) into Eq.~(\ref{eq:la}). This corresponds to
the limit $p_{F}\ll 2\varkappa \ll 1/d$ [Eq.~(\ref{eq:smalld})],
 where the condition $p_{F}\ll 2\varkappa $ is  automatically
 satisfied, since $r_s \sim 1$ and $2\varkappa /p_{F}=r_s N \gg 1$. The factor $(1+\cos \theta )/2$
entering Eq.~(\ref{eq:la}) is  a consequence of chirality. It
suppresses  backscattering and reduces $\lambda _{c}^{\text{max}}$
by a factor $2$ compared to $\lambda ^{\text{max}}$
[Eq.~(\ref{eq:lamax})], see Eq.~(\ref{eq:lamaxBCS}).

The obtained small value of $\lambda _{c}^{\text{max}}=1/16$
justifies the very applicability of the weak-coupling BCS approach
to determining $T_c$, within which the logarithm $\ln (\epsilon
_{F}/T)\approx 16$ has to be large. Therefore, the critical
temperature $T_c$ does {\em exponentially} depend on the inverse
coupling constant $1/\la_c$, which leads to its extremely small
value [Eq.~(\ref{eq:Tcmax})].

\section{Discussion and conclusion}
Let us now discuss the obtained results. Remarkably enough, as
Eqs.~(\ref{eq:lamax})-(\ref{eq:Tcmax}) demonstrate, the specifics
of the graphene spectrum (chirality and valley degrees of freedom)
appears to be very unfavorable for the realization of EC in
graphene-based devices. At the same time, this is not so for
double-layer systems based on materials with ``conventional''
metallic spectrum, such as, e.g., GaAs/Al$_{x}$Ga$_{1-x}$As
heterostructures used so far
experimentally~\cite{exp1,exp2,exp3,exp4}. Indeed, in such systems
the maximum interaction strength $\lambda _{c}^{\text{max}%
}=1/(N_{s}\,N_{l})=1/4$ is four times larger than in graphene
[Eq.~(\ref{eq:lamaxBCS})] due to the absence of the valley space
and chirality. This value is not that small and the system could
be on the verge of the weak-coupling limit. Therefore, one does
not get such a small value of the transition temperature as we
obtained for graphene.

It is also instructive to mention for comparison that in a
single-layer graphene subject to the in-plane magnetic
field~\cite{AKT}
one obtains  $\lambda _{c}^{\text{max}%
}=1/(2\,N_{s}\,N_{v})=1/8$ for the interaction constant, since the
system consists of only one layer ($N_{l}=1$) [see Eq.~(5.29b) in
Ref.~\cite{AKT}], and the exponential factor
$\exp (-\lambda _{c}^{\text{max}})=\exp (-8)\approx 3\cdot 10^{-4}$ in Eq.~(%
\ref{eq:Tcmax}) is not as small. However, the Zeeman splitting
energy enters Eq.~(\ref{eq:Tcmax}) instead of $\epsilon _{F}$,
which cannot be extremely high even for experimentally very high
magnetic fields $B$. For $B\approx 40\,\text{T}$ one can estimate
$T_{c}^{\text{max}}\sim 20 \text{mK}$.

\begin{figure}[tbp]
\includegraphics[width=.32\textwidth]{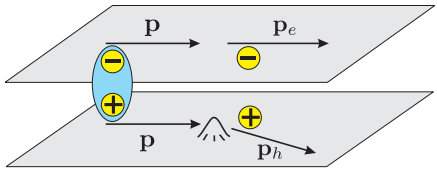}
\caption{ Sensitivity of the excitonic condensate to the impurity
scattering. Impurities with the size of potential smaller than the
interlayer distance scatter electrons and holes not identically,
thereby breaking electron-hole pairs and suppressing the
condensate.} \label{fig:scattering}
\end{figure}

There is another factor that is unfavorable for the
realization of EC in double-layer systems. Namely, %
the excitonic condensate is sensitive to the impurity
scattering~\cite{Zittartz,LY}. Since the bound electron and hole
carry the same momentum $ \mathbf{p}$, any scattering process that
changes the momentum of electron and hole not identically, i.e.,
$\mathbf{p}\rightarrow \mathbf{p}_{e}$ for electron and
$\mathbf{p}\rightarrow \mathbf{p}_{h}$ for hole, so that
$\mathbf{p}_{e}\neq \mathbf{p}_{h}$, breaks the electron-hole pair
(see Fig.~\ref{fig:scattering}). This is the case for any
impurities with the range of the scattering potential less than
the interlayer distance $d$, since the potential of such
impurities differs in the two layers. The effect of the impurity
scattering on the excitonic condensate
was studied analytically for conventional systems in Refs.~\cite%
{Zittartz,LY} and the theory is analogous to Abrikosov-Gorkov's
theory for magnetic impurities in superconductors. This approach
has been very recently applied to graphene in
Ref.~\cite{MDdisorder}. The main result of this study is that
sufficiently short-range impurities with the scattering time $\tau
$ destroy the excitonic condensate completely as soon as
\begin{equation}
\hbar /\tau \gtrsim T_{c},  \label{eq:disorder}
\end{equation}%
where $T_c$ is the transition temperature of the ideally clean
system.
Equivalently, for the condensate to exist, electron momentum has
to be conserved at the scale of the correlation length $\hbar
v/T_c$. Since the mean free path $v \tau \sim 1 \mu \text{m}$ of
the order of the typical size of graphene samples corresponds to
$\hbar/\tau \sim 1 \text{K}$, considering the values
$T_c^\text{max} \sim 1 \text{mK}$ obtained, the condensate should
be completely suppressed at any temperature $T\leq T_c^\text{max}$
even in the ballistic samples due to the boundary scattering.

In conclusion, we have studied the possibility of the excitonic
condensation in double-layer graphene systems. We have
demonstrated that in order to properly determine the transition
temperature, it is essential to take the screening of the coupling
interlayer Coulomb interaction into account. The specifics of the
graphene spectrum (chirality and valley degrees of freedom) leads
to a smaller interaction strength than in conventional
semiconductors and to an extremely small value $\lesssim
1\,\text{mK}$ of the transition temperature. This makes
graphene-based systems disadvantageous for the observation of the
excitonic condensation.

After the present work was completed, we became aware of
Ref.~\cite{Lozovikgraphene}, in which Eqs.~(\ref{eq:Tc}) and
(\ref{eq:la}) were obtained. However, contrary to our main
argument, that these equations are valid and provide good
quantitative prediction also for moderate to strong Coulomb
interactions ($r_s \gtrsim 1$), it was stated in
Ref.~\cite{Lozovikgraphene} that they should apply in the weak
coupling limit ($r_s \ll 1$ or $ p_F d \gg 1$) only, whereas for
$r_s \sim 1$, analogously to Refs.~\cite{MD2,ZJ}, one could expect
high $T_c$ of the order of room temperatures.

We thank Anatoly F. Volkov and Sergey V. Syzranov for useful
discussions. Financial support of SFB Transregio 12 is greatly
appreciated.

\end{document}